\documentclass{article}
\usepackage[utf8]{inputenc}

\usepackage{hyperref}
\usepackage{url}

\usepackage{booktabs}
\usepackage{amsfonts}
\usepackage{amsmath}
\usepackage{mathtools}

\usepackage{algorithm}
\usepackage[noend]{algpseudocode}

\usepackage{multicol}
\usepackage[letterpaper, margin=1in]{geometry}

\usepackage{titling}
\usepackage{authblk}

\usepackage{graphicx}
\usepackage[hypcap=false]{caption}
\newenvironment{myfig}{\medskip\begin{center}}{\end{center}\medskip}

\usepackage[square,numbers]{natbib}
\usepackage{graphicx}
\graphicspath{{./images/}}

\hypersetup{
    colorlinks=true,
    linkcolor=blue,
    filecolor=magenta,      
    urlcolor=cyan,
    citecolor=cyan,
}

\title{Modeling Image Quantization Tradeoffs for Optimal Compression}
\author{Johnathan Chiu}
\affil{University of California, Berkeley}
\date{November 2021}

\begin{document}
  \maketitle
  
\section*{\centering{Abstract}}
All Lossy compression algorithms employ similar compression schemes -- frequency domain transform followed by quantization and lossless encoding schemes. They target tradeoffs by quantizating high frequency data to increase compression rates which come at the cost of higher image distortion. We propose a new method of optimizing quantization tables using Deep Learning and a minimax loss function that more accurately measures the tradeoffs between rate and distortion parameters (RD) than previous methods. We design a convolutional neural network (CNN) that learns a mapping between image blocks and quantization tables in an unsupervised manner. By processing images across all channels at once, we can achieve stronger performance by also measuring tradeoffs in information loss between different channels. We initially target optimization on JPEG images but feel that this can be expanded to any lossy compressor.
 
\section{Introduction}
In this paper, we attempt to produce optimal quantization tables for lossy compression algorithms, most notably, JPEG. JPEG is a well known standard that has been widely used for many decades. Additionally, experiments are far easier to run given the algorithm is more simplistic and is more widely implemented than newer codecs like JPEG2000, HEIF, AVIF. The concepts still remain the same for RD optimization across all algorithms. 

We can formulate our proposed problem as an optimization one such that we want to maximize visual quality while minimizing file size. Given that this is evidently a non-convex problem, we can rely on neural networks to learn how to produce quantization tables which are locally optimal but still achieve the results that we desire. Additionally, rather than learning to linearly scale the table, we want to focus on each individual quantization index and how it relates to reconstructing the image as a whole. To train the model, we design a loss function that enables unsupervised learning so that our training process becomes the optimization process for each input.

Finally, we strive to achieve small file sizes while minimizing visual degradation. In this paper, we regard visual degradation by a number of parameters -- image "blockiness", significant color distortion, reduction of image texture. 

In this paper, we mainly reference Google's Guetzli \cite{GUETZLI}, MozJPEG \cite{mozjpeg}, and Kraken.io \cite{nekkraug} as our baseline metric for optimizers. 

\section{JPEG Basics}
The JPEG algorithm uses the Discrete Cosine Transform (DCT) followed by quantization to compress images \cite{DCT}. The quantization portion of the compressor is what makes JPEG a lossy compressor. In order to best compress the image, JPEG employs the use of run-length encoding (RLE) followed by Huffman encoding. In specific, the run length encoder is a zero-based run length encoder meaning that the data is compressed in a sequence of tuples where each tuple tells us the number of preceding zeros followed by the next succeeding value. Given the energy compaction properties of DCT, low frequency data and high frequency data are separated in the transformed matrix. A zig-zag encoding scheme helps associate the values from high to low-frequency in order. Additionally, previous knowledge about human visual perception define that higher frequency details in images are less perceptible to the human eye. The JPEG algorithm exploits the preceding two properties to heavily quantize higher frequency data in an image. 

\section{Related Works}
Research on RD optimization is already well characterized. Additionally, a number of state of the art algorithms were introduced many decades ago \cite{RDOPT, 334973, DCTune}. Many popular libraries such as MozJPEG still employ use of these specific works \cite{mozilla}. On the other hand, research in mathematical modeling of the Human Visual System (HVS) is still a large ongoing research topic \cite{sadykova2017quality, pop00031, hwang_chandak_tatwawadi_weissman_2021, 8954560}. 

The researchers at Google developed an algorithm to encode JPEG images using a closed-loop optimizer that utilizes Butteraugli, their proprietary model of human vision \cite{butteraugli}. The developers designed their model to make use of two major optimizations. The first involves making quantization tables coarser which decrease the magnitude of stored coefficients. The second involves directly replacing DCT coefficients in each block to optimize zero run-length encoding.

Likewise, MozJPEG targets the problem similar with Trellis Quantization and adaptive quantization. They optimize the global quantization tables using a Dynamic Programming Algorithm introduced by Crouse et. al \cite{551698}. 

The researchers in \cite{SemDCNN} developed a CNN that generates a map that highlights "semantically-salient" regions in a given image to encode them at a higher quality relative to the rest of the image. They claim the neural network improves compression by using higher bit rates to encode regions more sensitive to distortion according to the human perspective and lower bit rates elsewhere.

The authors of DeepN-JPEG made strong attempts to classify the impact of frequency bands in an 8x8 frequency block for the decompression of the JPEG algorithm \cite{liu2018deepnjpeg}. Zhao et. al introduced a loss function that proves to have better performace by combining the use of SSIM and $l_1$ loss \cite{7797130}. Both of the aforementioned papers were large inspiration in our work for modeling our loss function that provides higher quality information for RD optimization.

\section{Loss Function}
\subsection{Multi-scale Structural Similarity Index (MS-SSIM)}
Our loss function uses MS-SSIM as a means to measure image distortion. MS-SSIM incorporates use of the regular SSIM metric and low-pass filters up to a scale of $M$. Each individual low-pass filtered image is considered in the final formula. SSIM is a stronger metric than the traditional Mean-Squared Error (MSE) metrics since SSIM targets smaller local changes rather than the entire image. MS-SSIM between two images, $x, y$, is represented by the following formula(s):

\begin{equation}
\text{MS-SSIM}(x, y) = \text{SSIM}(x^M, y^M)^{\gamma_M}\Pi_{i = 1}^{M - 1}\text{CS}(x^i, y^i)^{\gamma_i}
\end{equation}

\noindent where 

\begin{equation}
\text{SSIM}(x, y) = \frac{2\mu_x\mu_y+C_1}{\mu_x^2\mu_y^2+C_1}\text{CS(x, y)}
\end{equation}

\noindent and 

\begin{equation}
\text{CS}(x, y) = \frac{2\sigma_{xy}+C_2}{\sigma_x^2+\sigma_y^2 + C_2}
\end{equation}

The following variables: $\sigma$ and $\mu$ follow their traditional representation of standard deviation and mean. $C_1$ and $C_2$ are arbitrary constants, usually 0.01 and 0.03 respectively. $\gamma$ is an arbitrary constant that weights the importance of each channel.

In addition, MS-SSIM takes into account structural information using the standard deviation of blocks in local regions.
Without going into significant technical information we invite interested readers to the original paper for more details \cite{MSSSIM}.

\subsection{Entropy Estimation}
Computing entropy is a nondifferentiable function since it requires the probability distribution of pixel values. We estimate entropy similar to the work in introduced by Ratnakar et. al \cite{RDOPT}. We make the slight modification to average the value across the samples at each coefficient index. In addition, we scale each coefficient index after averaging with a value denoted as $\alpha_i$ inspired by the work from Liu et. al \cite{liu2018deepnjpeg}. Our entropy estimation, $R$, over the quantized coefficients, $Q$, is thus defined as:

\begin{equation}
R(Q) \approx \frac{1}{N} \sum_{i = 0}^{63} \sum_{n = 0}^{N - 1} \alpha_i Q_n[i]
\end{equation}

where $N$ is the number of sample blocks taken from the image. Refer to section \ref{sec: NN} for more information on how $N$ is chosen.

\subsection{Minimax Function}\label{sec:minimax}
With the following information, we aim to minimize both our rate and distortion loss simultaneously. We incorporate the use of $l_1$ loss similar to work introduced by Zhao et. al \cite{7797130}. Thus, we can define our loss function with the following formula:

\begin{equation}
l(x, y) = -\beta \text{log}(\mathcal{L}^\text{MS-SSIM}(x, y)) + (1 - \beta) \mathcal{L}^{l_1}(x, y) + \gamma R(Q)
\end{equation}

$\beta$ and $\gamma$ are all hyperparameters that can be tuned. We additionally use a window size of 3 for our MS-SSIM function.

\section{Neural Network}\label{sec: NN}
\subsection{Network Input}
Our input consists of some number of samples of blocks in an image. The number of samples is a hyperparameter which can affect the fidelity of the image. We take inspiration from Prakash et. al but feel that it is simpler to find regions of interest (ROI) through simple statistical methods -- we sample more blocks with high variance  \cite{SemDCNN}. We also incorporate some randomness in sampling so we get a stronger spread across the image. The reasoning for this stems from the idea that high variance blocks can usually contain the most significant details in the image. So, if the quality following compression on high detail blocks are great, we can expect the rest of the image to have relatively high quality as well.

\subsection{Model Architecture}
\begin{figure}
    \centering
    \includegraphics[width=15cm]{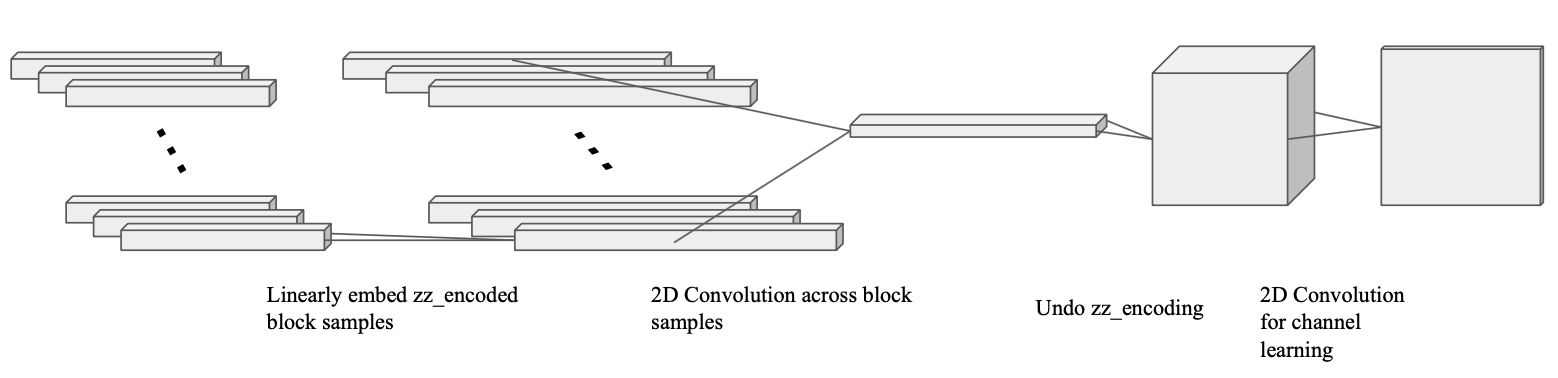}
    \caption{High level model architecture diagram}
    \label{fig:model_diagram}
\end{figure}

Our network incorporates the use of fully connected and convolutional layers. The fully connected layers are important in relating each frequency index to another in the zigzag-encoded data. The convolutional layers provide information across channels. We illustrate information on the architecture in diagram \ref{fig:model_diagram}. We start by zigzag encoding the input for each $C\times8\times8$ sample block, where $C$ is the number of channels. We similarly denote the number of sample blocks by $S$. We then linearly embed the $S\times C\times64$ zigzag encoded tensor into a $S\times C \times256$ tensor. This is followed by a 1D convolution which simultaneously downsamples the tensors and reduces the samples to 1 with a resulting shape of $1\times C\times64$. From here we undo our zigzag encoding and retrieve our $1\times C\times8\times8$ block and pass this into another series of convolution layers which gives us the quantization tables with shape $1\times Q\times8\times8$ where $Q$ is the number of quantization tables we desire.

\subsection{Training}\label{sec:training}
We use Deep Learning as a means to optimize over a singular image. Deep Learning works better than traditional optimization techniques. In particular, our model has the ability to learn across the three channels simultaneously and maximize rate by considering trade-offs in separate channels. The model is simply used to optimize and not as a means to learn information for any large number of images. We will attempt to extend this to a universal model trained on larger datasets. We discuss this point further in section \ref{sec:future_works}. 

In our training process, each image is optimized over 100 epochs. In addition, we use the Adam Optimizer provided in the PyTorch library to solve our function.

Despite our loss function being a strong model, we have no exact measurement to truly compare the tradeoffs between visual quality and compression rate. To find the best possible image, we require a bit of human supervision. We create bins for each image to fall into. These bins are split by MS-SSIM and we take each image with the lowest rate loss across all the bins. We then allow the user to choose the smallest image file with quality they find acceptable.

\subsection{Annealing}
As mentioned in \ref{sec:minimax}, $\beta$ and $\gamma$ are tunable hyperparameters. Rather than setting these constants, we look to use annealing as a means for enabling convergence. We penalize the $\beta$ term dependent on the MS-SSIM factor. When our MS-SSIM is small (more distortion), we penalize $\beta$ quite heavily. Otherwise, when we find that our model begins to converge to a sufficient MS-SSIM, we penalize $\beta$ by $ 1 - \text{MS-SSIM(x, y)} $ multiplied by some "temperature". And, we penalize $\gamma$ using the ratios of the entropy for the current solution over original solution also multiplied by the similar "temperature" value. The "temperature" is scaled each epoch in which the model proposes a solution that has an acceptable MS-SSIM value.

\section{Results}
We tested our model amongst a small set of 10 images with very differing qualities and saw really strong results throughout. We state the results of compression rates along with the respective resulting MS-SSIM below in figures \ref{fig:rate} and \ref{fig:dist}. 

\begin{myfig}
    \centering
    \includegraphics[width=15cm, height=5.5cm]{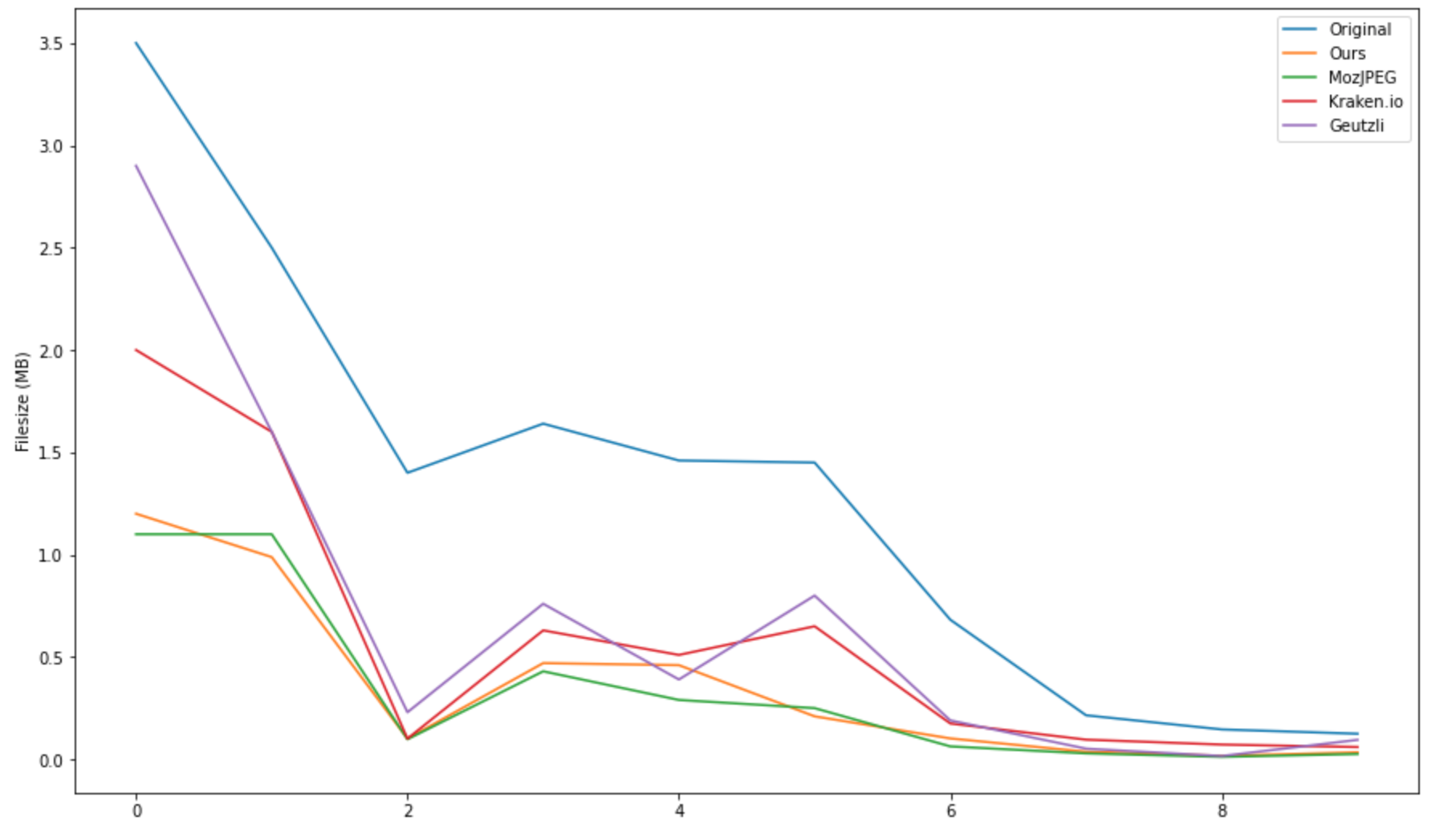}
    \captionof{figure}{Filesizes across test images}
    \label{fig:rate}
\end{myfig}

\begin{myfig}
    \centering
    \includegraphics[width=15cm, height=5.5cm]{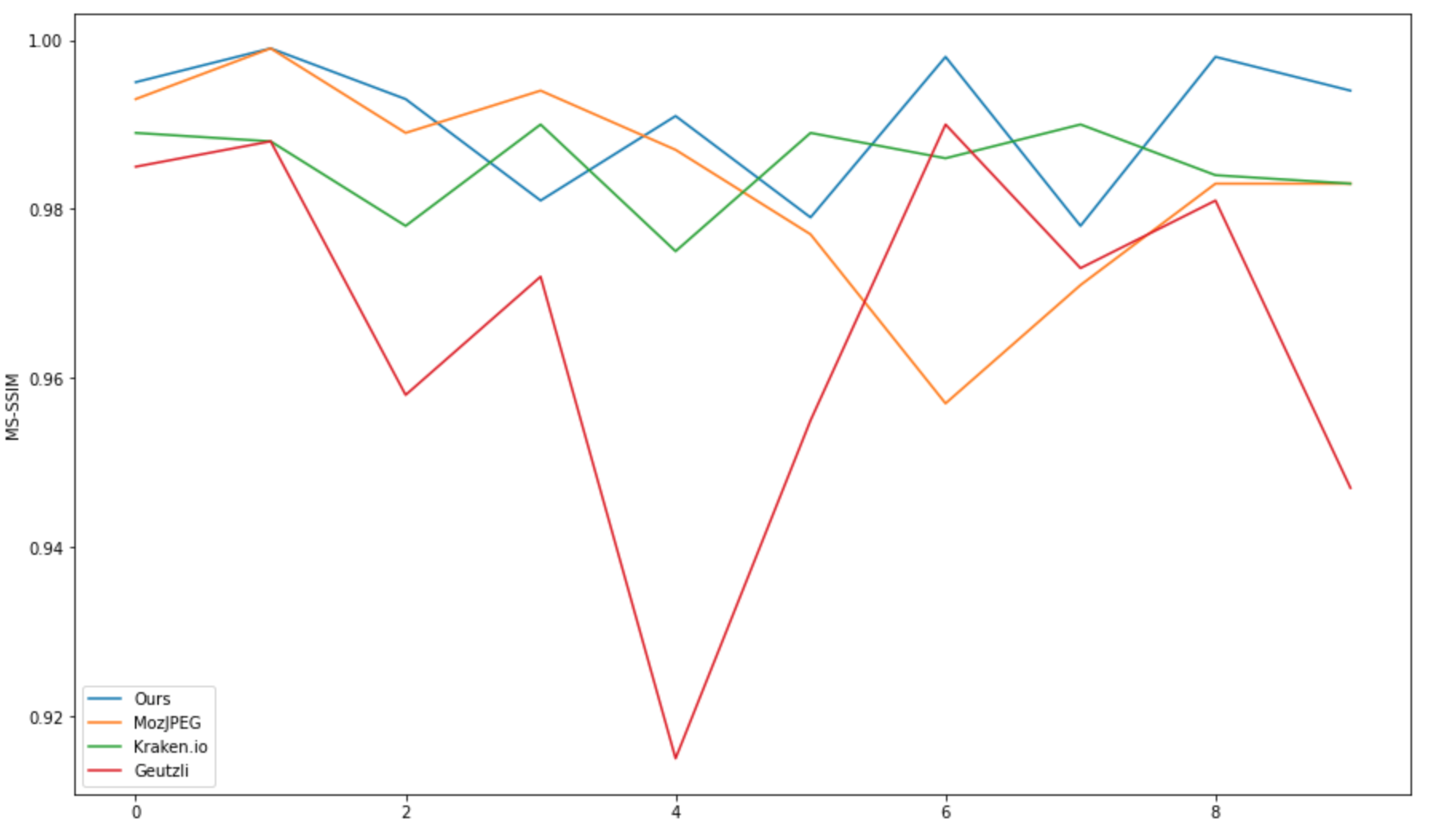}
    \captionof{figure}{MS-SSIM across test images}
    \label{fig:dist}
\end{myfig}

Our test set differed in a number of ways: (1) file size, (2) file resolution, (3) subjects, (4) high contrasting colors, and (5) number of channels. We found that our compressor had stronger ability to reduce distortion than the other compressors. We did also find that our algorithm also performed on par with MozJPEG despite their implementations of Trellis Quantization and adaptive thresholding. We include a small number of image examples in the Appendix for comparison.

\section{Future Work}\label{sec:future_works}

Another set of experiments that could show a great deal of promise would be to optimize a significantly larger model to learn and generalize outputs for any image input. This would enable the ability for these models to be implemented in a realworld setting.

The developers of MozJPEG used a number of adaptive thresholding techniques to produce smaller filesizes. We believe that implementing similar techniques will reduce file size exponentially and give us much stronger results. Given that our model alone gives really strong global quantization tables, we believe we can reapply the model on individual $8\times 8$ blocks to give strong local quantization tables that will reduce filesize without compromising image quality.

In section \ref{sec:training}, we note that we require human supervision to choose the most accurate output. We hope that future works can expand on HVS modeling and enable a more complete picture of human perception. We also found that a number of machine learning tasks face similar issues with alignment to more human-like understanding of scenes. With this work, our models will only improve and learn more human-like vision qualities. We would thus like to explore our loss function applied to tasks that expand beyond compression and to other topics in the vision space.

\section{Conclusion}
In this paper, we utilized Deep Learning as a tool to compressing images according to the JPEG file format. We demonstrate the applicability of Deep Learning to image compression (and maybe other notable types of compression) and hope that future research extends on developing strong mathematical models for HVS. We believe there are a large number of implications that demonstrate how machine learning models could potentially learn to examine images in the same manner as humans.
It is important to note that our work can truly expand beyond JPEG given the similar inner-workings of all lossy image compressors. We have strong reason to believe similar models can be applied to any given compression algorithm to achieve better results. And, despite the recent advances in other separate image compressors such as HEIF and AVIF, the ubiquity of JPEG still stands. 

Finally, though our test set is small, we hope that interested readers will test our work through our open source code repository. We hope this paper provides inspiration to readers to continue similar work and find novel and groundbreaking approaches to applying machine learning to data compression.

\bibliography{references}

\newpage

\section*{Appendix}

\vspace{1cm}

\begin{myfig}
    \includegraphics[width=.24\textwidth]{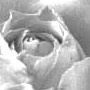}\hfill
    \includegraphics[width=.24\textwidth]{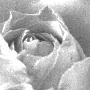}\hfill
    \includegraphics[width=.24\textwidth]{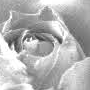}\hfill
    \includegraphics[width=.24\textwidth]{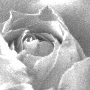}
    \captionof{figure}{\centering Test image 9 (original filesize: 146 kB), left to right: MozJPEG (12 kB), Ours (17 kB), Kraken.io (72 kB), Guetzli (16 kB)}
\end{myfig}

\begin{myfig}
    \centering
    \includegraphics[width=.24\textwidth]{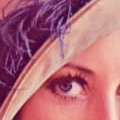}\hfill
    \includegraphics[width=.24\textwidth]{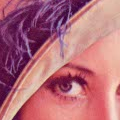}\hfill
    \includegraphics[width=.24\textwidth]{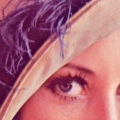}\hfill
    \includegraphics[width=.24\textwidth]{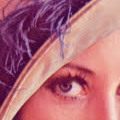}
    \captionof{figure}{\centering Test image 8 (original filesize: 215 kB), left to right: MozJPEG (29 kB), Ours (37 kB), Kraken.io (72 kB), Guetzli (52 kB)}
\end{myfig}

\begin{myfig}
    \centering
    \includegraphics[width=.24\textwidth]{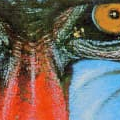}\hfill
    \includegraphics[width=.24\textwidth]{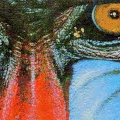}\hfill
    \includegraphics[width=.24\textwidth]{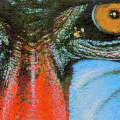}\hfill
    \includegraphics[width=.24\textwidth]{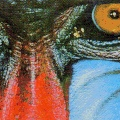}
    \captionof{figure}{\centering Test image 7 file (original filesize: 682 kB), left to right: MozJPEG (63 kB), Ours (102 kB), Kraken.io (175 kB), Guetzli (190 kB)}
\end{myfig}

\newpage

\begin{myfig}
    \centering
    \includegraphics[width=.4\textwidth]{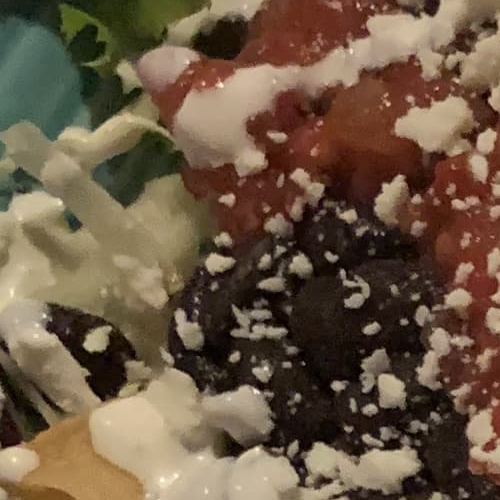}
    \includegraphics[width=.4\textwidth]{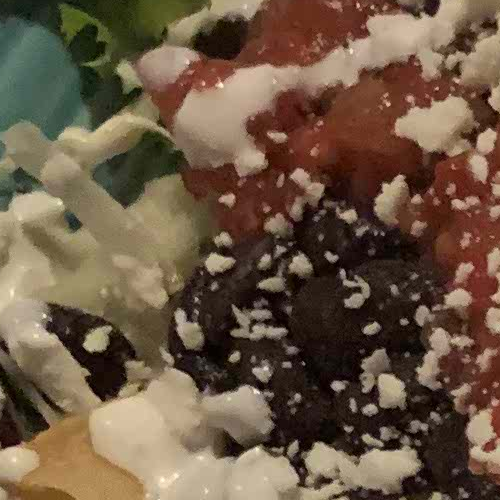}
    \captionof{figure}{\centering Not part of test set (original filesize: 3.7 MB), left to right: MozJPEG (1.1 MB), Ours (1.2 MB). The file dimensions were too large for Kraken.io nor Guetzli to generate}
\end{myfig}

\end{document}